\documentclass[sigconf]{acmart}

\AtBeginDocument{%
  \providecommand\BibTeX{{%
    \normalfont B\kern-0.5em{\scshape i\kern-0.25em b}\kern-0.8em\TeX}}}
    
\setcopyright{acmcopyright}
\copyrightyear{2022}
\acmYear{2022}
\setcopyright{rightsretained}
\acmConference[FAccT '22]{2022 ACM Conference on Fairness, Accountability, and Transparency}{June 21--24, 2022}{Seoul, Republic of Korea}
\acmBooktitle{2022 ACM Conference on Fairness, Accountability, and Transparency (FAccT '22), June 21--24, 2022, Seoul, Republic of Korea}\acmDOI{10.1145/3531146.3533113}
\acmISBN{978-1-4503-9352-2/22/06}

\usepackage{subcaption}
\usepackage{caption}
\usepackage{xcolor}
\usepackage[a-1b]{pdfx}

\usepackage{hyperref}
\hypersetup{final}

\begin{document}

\title{Exploring How Machine Learning Practitioners (Try To) Use Fairness Toolkits}

\author{Wesley Hanwen Deng}
\email{hanwend@cs.cmu.edu}
\affiliation{%
  \institution{Carnegie Mellon University}
  \streetaddress{5000 Forbes Ave}
  \city{Pittsburgh}
  \state{PA}
  \postcode{15213}
  \country{USA}
}

\author{Manish Nagireddy}
\email{mnagired@andrew.cmu.edu}
\affiliation{%
  \institution{Carnegie Mellon University}
  \streetaddress{5000 Forbes Ave}
  \city{Pittsburgh}
  \state{PA}
  \postcode{15213}
  \country{USA}
}

\author{Michelle Seng Ah Lee}
\email{michelle.sengah.lee@cl.cam.ac.uk}
\affiliation{%
  \institution{University of Cambridge}
    \streetaddress{Computer Laboratory}
  \city{Cambridge}
  \postcode{CB3 0FD}
  \country{UK}
  }

\author{Jatinder Singh}
\email{jatinder.singh@cl.cam.ac.uk}
\affiliation{%
  \institution{University of Cambridge}
    \streetaddress{Computer Laboratory}
  \city{Cambridge}
  \postcode{CB3 0FD}
  \country{UK}
  }

\author{Zhiwei Steven Wu}
\email{zstevenwu@cmu.edu}
\affiliation{%
  \institution{Carnegie Mellon University}
  \streetaddress{5000 Forbes Ave}
  \city{Pittsburgh}
  \state{PA}
  \postcode{15213}
  \country{USA}
}

\author{Kenneth Holstein}
\email{kjholste@cs.cmu.edu}
\affiliation{%
  \institution{Carnegie Mellon University}
  \streetaddress{5000 Forbes Ave}
  \city{Pittsburgh}
  \state{PA}
  \postcode{15213}
  \country{USA}
}

\author{Haiyi Zhu}
\email{haiyiz@cs.cmu.edu}
\affiliation{%
  \institution{Carnegie Mellon University}
  \streetaddress{5000 Forbes Ave}
  \city{Pittsburgh}
  \state{PA}
  \postcode{15213}
  \country{USA}
}

\renewcommand{\shortauthors}{Deng et al.}

\begin{abstract}

Recent years have seen the development of many open-source ML fairness toolkits aimed at helping ML practitioners assess and address unfairness in their systems. However, there has been little research investigating how ML practitioners actually use these toolkits in practice. In this paper, we conducted the \textit{first in-depth empirical exploration} of how industry practitioners (try to) work with existing fairness toolkits. In particular, we conducted think-aloud interviews to understand how participants learn about and use fairness toolkits, and explored the generality of our findings through an anonymous online survey. We identified several opportunities for fairness toolkits to better address practitioner needs and scaffold them in using toolkits effectively and responsibly. Based on these findings, we highlight implications for the design of future open-source fairness toolkits that can support practitioners in better contextualizing, communicating, and collaborating around ML fairness efforts.

\end{abstract}


\maketitle

\section{Introduction}

The real-world impacts of machine learning (ML) systems are rapidly expanding, influencing outcomes in education \cite{bosch2016detecting, Holstein}, healthcare \cite{esteva2017dermatologist, Sirinukunwattana2016LocalitySD}, credit scoring \cite{vigdor2019apple}, social media \cite{alvarado2018towards, eslami2015always}, public services \cite{De-Arteaga2020, Green2019, HoltenMoller2020}, and criminal justice \cite{feller2016computer, chouldechova2017fair}, among many other areas. A growing body of research has drawn attention to the ways these systems can, whether inadvertently or intentionally, serve to amplify existing social inequities or create new ones \cite{bolukbasi2016man, hardt2016equality, pmlr-v81-buolamwini18a, bender2021dangers, koenecke2020racial, obermeyer2019dissecting, asplund2020auditing}. In response, recent years have seen the development of many open-source ML "fairness toolkits" intended to assist ML practitioners in assessing and addressing unfairness in the ML systems they develop \cite{bird2020fairlearn, bellamy2018ai, saleiro2018aequitas, bantilan2018themis, wexler2019if, adebayo2016fairml, tramer2017fairtest}. For instance, companies such as Microsoft, Google, and IBM, have published combinations of toolkits and guidelines \cite{Amershi2019GuidelinesFH, ghosh2021uncertainty, googlePAIR, arnold2019factsheets} that incorporate fairness as part of their core values.

Despite growth in the development and dissemination of fairness toolkits, there has been little research investigating how ML practitioners \emph{actually use} these toolkits in practice. In order to explore practitioners' perceptions and desires around open-source fairness toolkits, Lee et al. conducted interview studies and a survey to identify the gaps between the capabilities of existing fairness toolkits and the needs of industry practitioners \cite{lee2020landscape}. In a similar vein, Richardson et al. conducted an interview study with twenty ML practitioners in a simulated scenario in order to generate a practitioner-oriented rubric for evaluating fair ML toolkits \cite{richardson2021towards}. However, neither of these two works engaged practitioners \textit{directly} in using a fairness toolkit within the context of a real ML task. Prior literature suggests that the design of fairness toolkits that practitioners will find usable and useful in practice is a complex problem. For example, ML practitioners often find it challenging to appropriately \textit{formulate the problem} when translating a real-world fairness question into a form amenable to quantitative fairness assessment \cite{passi2017data, holstein2019improving, passi2018trust, zhang-2020-datascience}. When faced with a fairness-related challenge, no single developer is likely to have all of the cultural and domain knowledge relevant to understanding or addressing the issue \cite{holstein2019improving, sambasivan2021re, andrus2021we}, and the appropriateness of different fairness definitions may be socially contested \cite{passi2019problem, mulligan2019thing, mitchell2018prediction, wang2020factors}. Adding an additional layer of complexity, the contextual nature of ML fairness \cite{Dasch2020OpportunitiesFA, van2021effect, 10.1145/3178876.3186138, lee2021formalising, lee2021bias} makes it particularly difficult to design general-purpose toolkits that can effectively support practitioners in assessing and addressing fairness across a wide range of ML applications and real-world contexts. To better understand and improve the usefulness and usability of software toolkits, prior research in Human-Computer Interaction (HCI) has emphasized the importance of studying how practitioners actually attempt to use toolkits in the context of real-world tasks \cite{Murphy2018APIDI, Myers2016ProgrammersAU}. One effective approach is conducting "think aloud" interview studies, in which participants are asked to continuously articulate their thinking while exploring and using a software toolkit \cite{Myers2016ImprovingAU, sushine2015searching, stylos2008implications, wijayarathna2019empirical, ellis2007factory}.

In this study, we conduct think-aloud interviews and a survey to explore the following two questions: \textit{1. How do practitioners (try to) work with existing ML fairness toolkits? 2. What opportunities exist for fairness toolkits to better support them during these phases?} We first designed a realistic ML task in which we required practitioners to build an ML model based on a real-world dataset to help allocate education resources, while thinking about the potential fairness issues present in the dataset and their model. After screening forty-one industry practitioners who responded to our recruitment survey, we invited twenty-three ML industrial practitioners to undertake the task before joining the interview study. In the end, eleven practitioners finished the ML task and completed our think-aloud interview study, in which they encountered two fairness toolkits \emph{for the first time}, explored the toolkit APIs, and tried to use them to address fairness issues in the ML models they built, while "thinking aloud" their observations, thoughts, and confusions. We also conducted an anonymous online survey with fifty-six industry practitioners who encountered fairness issues and might have used fairness toolkits %space saving change...
before, so as to further explore and supplement our interview observations.

Through our investigation, we found that practitioners desire better support from fairness toolkits to better contextualize ML fairness issues and help communicate often complex fairness analysis to non-technical colleagues in their work places. We also identified four distinct design requirements \cite{Murphy2018APIDI, Myers2016ImprovingAU} ML practitioners had when using fairness toolkits, namely, (1) the abilities to use the toolkit to learn more about ML fairness and the landscape of current ML fairness research, (2) rapidly on-boarding toolkits due to workplace time constraints, (3) the abilities to integrate the toolkits into existing ML working pipeline, and (4) using toolkits as code repositories to implement state-of-the-art or domain specific ML fairness algorithms. In addition, we surface contexts in which practitioners committed to pitfalls \cite{Murphy2018APIDI} while addressing fairness issues, as well as misused the toolkits largely due to organizational time constraints for fairness work. Informed by our findings, we highlight implications for designing future open-source fairness toolkits. More broadly, our work contributes to growing efforts from both academia and industry towards ensuring that advances in ML fairness research have positive impacts in practice.

\section{Background and Related Work}

\subsection{Understanding Practitioners' Needs and Challenges around ML Fairness}

A number of recent studies have investigated challenges that ML practitioners face when attempting to improve fairness in practice. For instance, through interviews and surveys with commercial product teams, Holstein et al. \cite{holstein2019improving} identified many disconnects between the solutions offered by the fair ML research literature (and toolkits implementing these solutions) versus the real-world challenges faced by industry ML practitioners. Through interviews and co-design workshops, Rakova et al. \cite{rakova2020responsible} and Madaio et al. \cite{madaio2020co, madaio2021assessing} investigated the organizational challenges, tensions, and barriers that practitioners face in practice when attempting to improve fairness in ML products and services.

More recently, in response to a surge of open source ML fairness toolkits (e.g., \cite{bird2020fairlearn, bellamy2018ai, saleiro2018aequitas, wexler2019if}), Lee et al. \cite{lee2020landscape} undertook a comparative assessment of the strengths and weakness of six prominent open source fairness toolkits and identified gaps between these toolkits' capabilities and practitioners' needs. In addition, Richardson et al. \cite{richardson2021towards} created a rubric containing two main evaluation criteria for fairness toolkits, through an interview in which practitioners reviewed analysis results generated by fairness toolkits, which were curated by researchers. While we build upon and contrast against findings from these two important prior studies in our current work, both of these studies relied on retrospective interviews and survey techniques to understand practitioners' challenges and their experiences with fairness toolkits. 
%\js{tweaked the next bit - old version in comments}
As a result, both studies offer valuable insights regarding the usability and design of ML fairness toolkits. However, yet to be considered are the challenges faced by practitioners when using fairness toolkits to perform a task. 

The current study represents the \textbf{first task-based exploration} in the literature %\js{do we need this change? clunky. } 
of how ML practitioners (try to) learn about and work with fairness toolkits and their APIs. 

\subsection{Open Source Fairness Toolkits}

ML open-source fairness toolkits intend to assist ML practitioners in assessing and (potentially) mitigating unfairness in the ML systems they develop. Fairness toolkits usually offer ready-to-use fairness metrics and mitigation algorithms \cite{Tutorial} as their main functionalities. In short, a \emph{fairness metric} is a quantification of unwanted bias in training data or models. A \emph{bias mitigation algorithm} is a procedure for reducing unwanted bias in training data or models. Some popular fairness toolkits include Fairlearn \cite{bird2020fairlearn}, AIF360 \cite{bellamy2018ai}, Aequitas \cite{saleiro2018aequitas}, Themis-ML \cite{bantilan2018themis}, What-If Tool \cite{wexler2019if}, Fair-ML \cite{adebayo2016fairml}, and Fair-Test \cite{tramer2017fairtest}. In our study, we investigated how practitioners worked with two toolkits that were identified by Lee and Singh \cite{lee2020landscape} as among the most useful and well-documented: IBM's AIF360 \cite{bellamy2018ai} and Microsoft's Fairlearn \cite{bird2020fairlearn}. Both toolkits contain a Application Programming Interface (API), i.e., an interface that practitioners work with in order to communicate with toolkits \cite{Myers2016ImprovingAU}. We briefly introduce these two toolkits below.

\subsubsection{AIF360}
Developed by IBM, AI Fairness 360
(AIF360)\footnote{https://aif360.mybluemix.net/} is an extensible open source toolkit for detecting, understanding, and mitigating algorithmic biases \cite{bellamy2018ai}. With over 71 bias detection metrics and 9 bias mitigation algorithms (suited for all aspects of the ML pipeline- from pre-processing to in-processing to post-processing), AIF360 is often commended for its breadth and depth in the coverage of fairness-related topics \cite{hufthammer2020bias, johnson2020fairkit, lee2020landscape}. Despite this, IBM notes that the toolkit should only be used in a very limited setting: allocation or risk assessment problems with well-defined protected attributes.  

AIF360 is also part of IBM's current effort towards ``trustworthy AI,'' an initiative focusing on creating a holistic approach to governed data and AI technology. Currently, AIF360 has 1,600 stars on GitHub and 514 forks as we are writing this paper. Also, there are 1,220 members in AIF360 slack channel, which is primarily used by engineers who ask for help regarding their syntax errors and other bugs while using the toolkit.

\subsubsection{Fairlearn}
Initially developed by Microsoft Research and now being maintained as a community-driven open source project, Fairlearn\footnote{https://fairlearn.org/} is an open source toolkit that empowers data scientists and developers to assess and improve the fairness of their AI systems that focuses on negative impacts—specifically, allocation harms and quality-of-service harms—for groups of people. The goal of Fairlearn is to create a vibrant community and resource center that provides not only code, but also resources such as domain-specific guides for when to use different fairness metrics and bias mitigation algorithms \cite{bird2020fairlearn}. To this end, Fairlearn boasts a detailed User Guide, which is meant to be complementary to their standard API documentation. Fairlearn has 1,200 stars on GitHub and 280 forks as we are writing this paper. The Fairlearn development team holds a weekly community call which practitioners can join through Discord, a chat and networking platform.

\section{Methods}

\subsection{Think-aloud Usability Evaluation Interview}

\subsubsection{Participants}

Before beginning our study with industrial practitioners, we first recruited two university computer science students and conducted semi-structured interviews as our pilot study to help us polish and iterate upon our study protocol. For our formal interview study, we recruited practitioners working on ML products and services through a combination of purposive and snowball sampling \cite{sampling}. We invited an initial set of participants through our personal connections with industry practitioners, and we then asked them to help disseminate our recruitment message through their networks. We also included a brief screening form to help us target suitable participants for our study. 
We specifically recruited ML practitioners who had previously encountered fairness issues in their professional work, and who had \textit{never} used a fairness toolkit prior to joining our study. Participants had on average five years of programming experience and three years of ML experience. This enabled us to observe how practitioners who were knowledgeable about real-world fairness challenges approached learning about fairness toolkits for the first time.

Overall, forty-one industry practitioners responded to our recruitment message. Twenty-three responded to our follow-up emails, among which fourteen participants scheduled an interview time. In the end, twelve participants attended the interview study, and eleven of them completed all phases of the study.  \hyperref[tab:freq]{Table 1.} provides details about participant demographics and their relevant experience. Specific details about their companies and working environment have been abstracted to preserve anonymity. All participants (P1 - P11, U1, U2) were compensated with a \$50 Amazon gift card upon completion of the study. The study was approved by our Institutional Review Board.

\begin{table*}
  \label{tab:freq}
  \begin{tabular}{ c c c c c} 
     \toprule
     Participant ID & Academic Background & Industrial Role \& Domain & Company Size & Location\\ 
     \midrule
     P1 &  Computer Engineering & Researcher, Tech Company & 50-249 & US \\ 
     P2 &  Computer Science & Data Scientist, Financial Services & 25,000 or more & UK \\
     P3 &  Computer Science & Machine Learning Engineer, Legal Tech & 50-249 & US \\
     P4 & N/A & Manager, Consulting Firm & 5,000 - 24,999 & Africa \\
     P5 &  Computer Science & Applied Scientist, Tech Company & 25,000 or more & US \\
     P6 & Human Rights & Researcher, Education & N/A & UK \\
     P7 & Sociology & Data Scientist, Retail & 5,000 - 24,999 & Sweden \\
     P8 & N/A & Data Scientist, Public Sector & 25,000 or more & UK \\
     P9 & Computer Science & Data Science Manager, Oil/Gas &  25,000 or more & US \\
     P10 & Business & Sr. Business Data Analyst, FinTech & 25,000 or more & US \\
     P11 &  Computer Science & Machine Learning Engineer, Legal Tech & 5,000 - 24,999 & US \\
     \hline
     U1 & CS, Undergraduate & N/A & N/A & US \\
     U2 & CS, Undergraduate & N/A & N/A & US \\
     \bottomrule
    \end{tabular}
    \caption{List of Pilot and User Study Participants.}
\end{table*}

\subsubsection{Study Design}

With the goal of observing how industry practitioners formulate and attempt to resolve a fairness-related problem with the aid of open-source fairness toolkits, we designed a task setting that required participants to engage with a complex real-world context. Specifically, participants were tasked with building a model to determine which students were in need of additional tutoring resources. For this study, we chose the Student Performance dataset \cite{Student}, which is concerned with academic achievement in Portuguese secondary education schools. Attributes include student grades as well as demographic, social, and school-related features \cite{cortez2008using}. Since we aimed to observe participants' thought processes during the Exploratory Data Analysis (EDA) and problem formulation stages, we desired a dataset that participants would be less familiar with. Hence, we intentionally avoided more commonly used datasets such as the COMPAS Recidivism Risk Score dataset \cite{angwin2016machine}. We also elected to utilize this dataset due to its reasonable size considering our study time (around 650 instances), as well as its inclusion of multiple features (both categorical and quantitative) which satisfy prevailing notions of sensitive attributes \cite{bird2020fairlearn, friedler2019comparative}. For example, the features concerning parent education and family educational support might prompt practitioners into investigating how socio-economic aspects might affect student performance and how one should treat these features while building models. These intricacies allowed us to delve deeper into the justifications behind choices that participants may make (e.g., selection of sensitive features).

\subsubsection{Interview Study Protocol}

The interview study consisted of a pre-interview task and a 60 minute think-aloud semi-structured interview. We now document our procedures for the full interview.

\emph{Pre-interview task:} Before entering the interview, we ask participants to complete a selection of preparatory tasks in the Colab notebooks (an collaborative computational notebook) we shared with them through email. Each participant received a notebook titled with a unique number string. After setting up the coding environment, we briefly described the Student Performance dataset and introduced practitioners to the task, in which they need to predict students' future school performance to help teachers distribute tutoring resources more efficiently. The entire pre-interview required practitioners to explore the data, translate the real-world problem into a machine learning one (problem formulation), train a machine learning model, and answer some questions regarding these steps in a pre-survey. 

The pre-interview task took 30 - 60 minutes to finish based on participants' self-reports. We share the Colab notebook we used  \href{https://colab.research.google.com/drive/1zc4V-KmY_VcAY2K2iMcgslK6j3Ip5m9x?usp=sharing}{\textit{through this link}} to help and inspire others conducting future relevant fairness toolkits evaluations.

\emph{Think-aloud semi-structured interview: } During the live interview, we began by taking 5-10 minutes and asked participants to elaborate on their responses to the aforementioned questions. Then, we spent the next 30 minutes letting them explore two of the most widely known fairness toolkits \cite{lee2020landscape}, namely Microsoft Research’s Fairlearn and IBM’s AI Fairness 360 (AIF360). After participants spent roughly equal amounts of time exploring the two toolkits, we asked them to select one of the two toolkits to directly implement their fairness assessment and mitigation code in Python within the Colab notebook. Throughout the interview, we also asked participants to provide feedback on whether they could envision a setting where they used toolkits in practice, and the various obstacles which might be present if toolkits are to be integrated into their daily workflow. Importantly, as participants traversed through the API interfaces and implementing codes, we encouraged participants to "think aloud" \cite{Someren1994TheTA,Kupis2019AssessingTU} and discuss the various information that was being displayed and how their understanding of the task (and ML fairness) was developing.

\subsubsection{Data Analysis}

We used an inductive thematic analysis approach \cite{braun2006using, clarke2014thematic} to analyze approximately 11.5 hours of video recordings and their corresponding transcripts.\footnote{The interview videos and audios were recorded in line with participant consent.} Two of the authors first worked together with a research assistant to conduct an open coding of the transcripts. We coded the same transcripts, discussed the code with the entire research team, then divided the rest of the transcripts and videos. We cross-compared and grouped these codes and observations into successively higher-level themes concerning the relationships between practitioners’ practices and fairness toolkits’ current functionalities and limitations. In Section \ref{4}, we discuss the findings identified from these codes and themes, together with implications for future fairness toolkits design. We share our initial round of thematic analysis 
\href{https://docs.google.com/spreadsheets/d/1l2hvAVCRSBtpqLzs_05sWgf7CAUxKZ0CtCBSK6AMnvk/edit?usp=sharing}{\textit{through this link}.}

\subsection{Anonymous Online Survey}

We then conducted an online survey of industry practitioners to better understand the real-life drivers and obstacles to using fairness toolkits, and also to supplement the interview findings with larger sample size. The survey contained three main sections: (i) background and information questions asking participants' industry domain and relevant experience in ML, ML fairness, fairness toolkits; (ii) questions about fairness toolkits themselves; and (iii) questions about using fairness toolkits (and fairness in general) in the practitioners' workplace. All questions were optional.

This online survey was anonymous and did not ask for any directly identifying information. We emailed the survey link to direct contacts, as well as advertising it on online communities related to ML and ML fairness, e.g. those on Reddit, LinkedIn and Slack channels. We also encouraged sharing of the survey link to relevant practitioners. Of the 71 people who started the survey, 56 (79\%) of the respondents completed at least one entire section, and 21(30\%) completed the entire survey. Therefore, the sample size varies with each question (between 21 and 56). A similar drop-out rate was encountered by Lee et al. in their prior anonymous survey study \cite{lee2020landscape}. Note that the questions would be difficult to contextualize for a respondent without a background in fairness-related challenges. This may have contributed to the drop-out rate, suggesting that few practitioners have relevant expertise or are not confident with issues of fairness. 
In addition, sensitive topics about organizational culture and workplace dynamics in the last survey section might also contribute to the drop-out rate. In spite of this, the opinions of this niche group are highly relevant. In addition, we were able to collect rich qualitative data from the free-text fields, in which many of the respondents discussed their experience with fairness toolkits. The anonymous survey data are available \href{https://docs.google.com/spreadsheets/d/1xo75vv3AAu4ADKLs_fYJJ2l79QbkEPgh/edit?usp=sharing&ouid=117517882265320108005&rtpof=true&sd=true}{\textit{through this link}.}

\section{Findings} \label{4}

We present findings from our think-aloud interview study, divided into three main phases: preparing to evaluate fairness; exploring and learning about toolkits; and attempting to use toolkits. Across all three phases, we found that practitioners desired greater support from toolkits in contextualizing, communicating, and collaborating around ML fairness efforts. We supplement observations from think-alouds with results from our online survey. At the end of each section, we share implications for the design of future fairness toolkits. 

\subsection{Preparing to Evaluate Fairness} 

During the Exploratory Data Analysis (EDA) and problem formulation phases of our study, we observed that practitioners drew heavily upon their personal experiences to surface potential sensitive features. However, they also recognized the limitations of their own knowledge and experience, expressing desires for help from domain experts in formulating relevant and coherent fairness-related questions for a given real-world context. When formulating their questions for fairness assessment, we observed that some participants appeared to be influenced by a toolkit’s specific functionalities and limitations.  We close this section by discussing implications of these observations for future toolkit designers.

\subsubsection{Participants’ analysis choices were heavily influenced by their personal experiences, knowledge, and beliefs.} \label{4.1.1}
EDA is a critical step in ML development \cite{polyzotis2018data, wang2021much} and an important opportunity to surface systematic errors and biases in datasets \cite{pmlr-v81-buolamwini18a, gebru2021datasheets, scheuerman2021datasets, datacascades}. We observed that during the EDA phase, participants often drew heavily upon personal experiences when making judgments about potentially sensitive features. For example, when explaining their concerns about the ``father's job'' and ``mother's job'' features, P5 said: \emph{``it was less from the machine learning perspective, but more from my personal experience as a teacher before... if the parents have higher education, it gives some cognitive bias to the teacher.''} We also observed that participants often relied upon general heuristics when deciding which features were potentially sensitive in a given domain. For example, nine out of eleven participants (P1, P2, P4, P5, P7, P6, P8, P9, P12) assumed that sex was a sensitive feature, believing this to be an obvious starting point. For example, as P8 said, \emph{“the first thing is, of course, check the sex.”}

Interestingly, most (seven out of nine) of the participants who assumed sex was a sensitive feature (P2, P4, P6, P7, P8, P9, P11) \textbf{attempted to mitigate biases in the ML pipeline by simply removing or ignoring the sensitive features like sex or address}. P9, for example, argued that \emph{``I feel that sex is one of the sensitive [features]. To make the model fair, I'd rather just remove it before training (the model).}'' This assumption has been discussed in prior literature as ``fairness through unawareness.'' \cite{dwork2012fairness} In reality, omitting sensitive features may lead to more disparate outcomes in practice. Furthermore, none of these seven participants considered whether seemingly neutral features might be a proxies for other sensitive attributes. 

\subsubsection{Some participants (re)formulated the ML problem to a format for which current fairness toolkits provide more support}

Through six months of ethnographic fieldwork with a corporate data science team, Passi et al. uncovered the "negotiated, not faithful, translation" of ML problem formulation which was affected by available tools and organizational resources \cite{passi2019problem}. In our study, we observed that \textbf{participants would sometimes reformulate problems based on current fairness toolkits functionalities and limitations}. For instance, after realizing that both toolkits offer better support for classification problems, five out of seven participants who had initially formulated regression problems during the pre-task phase decided to reformulate their problem as classification. For example, P2 emphasized that they would always prefer to use the example notebook toolkits offered as references whenever possible in their work. P2 then reformulated the ML problem from linear regression to \emph{DecisionTreeClassifier}, the classifier used in Fairlearn example notebook, since \emph{``it’s easier this way to use FairLearn's example notebook as supporting material.''} P4 switched to classification after realizing more comprehensive support for classification from both toolkits, adding that \emph{``I would have just framed a classification problem if I [knew] in advance that [the] toolkit supports that more.''} Survey participants expressed similar concerns through a free-text question about current toolkits limitations. For instance, one survey participant reported, \emph{``It was hard to apply AIF360 to many of our models which are not binary classification; e.g., for regression models, there is not much guidance on if or how the toolkit should be used.''} Only two participants in our think-aloud study (P5 and P9) raised concerns about reformulating the problem because of the toolkit's limitations. For example, P9 commented that \emph{``type of model or problem should not be dependent on the functionality of the toolkits.''}

\subsubsection{Participants were conscious about the need for expert guidance to inform analysis choices at the EDA phase} \label{4.1.3}

We observed that \textbf{participants were eager for guidance from domain experts and other relevant stakeholders at the EDA phase}. For instance, P3 mentioned that they wanted to consult dataset builders about \emph{``how the data was collected and how the features [were] being defined''}  P1, P8, P10, and P11 all wished to present their data analysis results to domain experts in order to understand \emph{``implications of technical concepts under different social context[s]''} (P1). For example, while explaining their EDA analysis results, P11 emphasized that they would \emph{``definitely chat with education experts and legal experts even before the modeling.''} 

When identifying potential sensitive features, P7 pointed out that \emph{``race is not common to have [as a dataset feature] in Sweden,''} noting that they would want to \emph{``consult with domain experts before deciding which features could potentially risk introducing bias.''} P4 mentioned the need to discuss \emph{``which features might generate what type of bias with domain experts in the Portuguese education system''}. From our survey, of the twenty-seven respondents who answered the question regarding which experts they would engage on a fairness issue, twenty-one (78\%) had \emph{``legal and regulatory experts''} as one of their selections. Twenty (74\%) selected \emph{``business domain experts,''} and fourteen (52\%) selected \emph{``reputational risk experts.''}

\subsubsection{\textbf{Implications}} \label{4.1.4}

\begin{itemize}
    
    \item \textbf{Broaden the scope of fairness toolkits to across the ML development lifecycle}. A large body of recent FAccT literature has highlighted the importance of exploring and analyzing datasets to surface potential biases \cite{pmlr-v81-buolamwini18a, gebru2021datasheets, scheuerman2021datasets, denton2021genealogy, paullada2020data, miceli2020between, datacascades}. However, current fairness toolkits offer little support for early stages of the ML development lifecycle, such as the EDA and problem formulation stages \cite{richardson2021towards, lee2021bias, cobbe2021reviewability, miceli2022studying, lee2021bias}. For instance, HCI researchers have explored the design of interactive interfaces like Facets \cite{facets}, FairVis \cite{cabrera2019fairvis} and FairSight \cite{ahn2019fairsight} to help users explore datasets and discover potential biases. To support the EDA phase, fairness toolkits should consider including similar interfaces in their designs. In addition, Boyd et al. found that context documents like \textit{Datasheets} \cite{gebru2021datasheets} could better scaffold an ML practitioner's process of issue discovery, understanding, and ethical decision-making around ML training datasets\cite{Boyd2021DatasheetsFD}. Future fairness toolkits could include instructions and educational materials on creating and reviewing \textit{Datasheets} \cite{gebru2021datasheets}, and similar documentations like "Dataset Nutrition Labels" \cite{Holland2018TheDN} and \textit{Model Cards}~\cite{mitchell2019model} to better inform practitioners' data explorations. 
    
    \item \textbf{Design fairness toolkits to facilitate interdisciplinary conversations and collaborations}. In our study, participants expressed desires for guidance from domain experts and other relevant stakeholders, to better understand the social and cultural context in which a given dataset or ML system is situated. Therefore, we suggest that future fairness toolkits might be explicitly designed as \emph{social computing systems} that help to facilitate conversations between different toolkit users with diverse backgrounds and knowledge (e.g,. by connecting different toolkit users for peer support on an ad hoc basis). For example, toolkit users with technical backgrounds but lack certain expertise in law or gender study could seek help from relevant domain experts through the potential social computing functions offered by the future open-sourced fairness toolkits. Fairness toolkits could also introduce interactive, deliberation-driven design activities to encourage critical reflections and facilitate interdisciplinary conversations around ML fairness issues (cf. \cite{shen2021value,wong2021timelines}). 

    \item \textbf{Show practitioners both patterns and anti-patterns for toolkit use}. Helping users recognize, diagnose, and recover from errors is essential to the design of usable APIs and toolkits \cite{Myers2016ImprovingAU}. In our study, a large proportion of participants committed to the “fairness through unawareness” trap \cite{dwork2012fairness}. Prior work suggests that timely, clear warnings and error messages can be effective in helping users avoid common pitfalls \cite{Myers2016ImprovingAU}. In the case of fairness toolkits, one possible design opportunity to help users avoid various "fairness traps" is to display contextual warning messages that target common pitfalls. In addition, in many domains like medicine \cite{Hales2006TheCT}, aviation \cite{Burian2006DesignGF}, and structural engineering~\cite{Forbes2010ModernC}, checklists are used to support task completion, guide decision making, and prompt critical conversations among stakeholders \cite{Lingard2005GettingTT,madaio2020co}. Well-designed fairness checklists may help practitioners avoid oversimplifying or forcing an ML problem formulation due to toolkits constraints \cite{madaio2020co}.

\end{itemize}

\subsection{Exploring and Learning about Toolkits}

Through the ``exploring and learning'' phase of our study, we identified four major design requirements for fairness toolkits among participants, discussed below. Based on our observations, we discuss future implications for the design of fairness toolkits that can serve the needs of a diverse group of potential users.

\subsubsection{Some participants wanted to be able to use toolkits as educational tools}. Instead of directly applying fairness toolkits to their current projects to address fairness issues, some participants were most interested in using toolkits to \textbf{learn more about ML fairness concepts and terminologies.} P2 said that for a fairness toolkit, \emph{``ML fairness is something new... not directly [related] to my work and I just got into it maybe a few months ago... the most important thing for me is the [explanations of metrics] instead of the mitigation [code].''} While going through both APIs, P6 commented that, in their future work, they would like to use AIF360 as a starting point to learn more about different fairness metric definitions and relevant academic papers, in addition to their use cases. P10 pointed out that, since \emph{``there is a lack of [resources] to learn more about fairness in [my] company''}, they appreciated that AIF360 offered a broad view of state-of-the-art ML fairness techniques, and included a designated reference section for convenience. Our survey results further supported this finding. For instance, one survey respondent commented on the lack of organizational process or \emph{``official training and awareness raising''} around fairness issues. In addition, for the question: ``What are some of the following options that most likely be your reason(s) to explore an open source fairness toolkit?,'' we had thirty-two out of thirty-nine (82\%) selected ``learn more about ML fairness concepts and terminologies'', and twenty-seven (69\%) selected ``learn more about the typical process of dealing with ML fairness.'' Out of thirty-one survey respondents for whom these questions applied, only five (16\%) said they have a defined process in their organisation for fairness, while fifteen (48\%) said they do not have such a process, and eleven (35\%) were not sure. Out of twenty-four survey respondents who specified whether their organization would provide sufficient time and resources to address a fairness-related concern, responses were nearly evenly split, with ten (48\%) responding ``No'', and eleven (52\%) responding ``Yes.'' The drop-off of thirteen respondents at this stage of the survey may be due to the sensitive nature of this question.

\subsubsection{Some participants wanted rapid onboarding to fairness toolkits due to workplace time constraints}. Some participants preferred to \textbf{learn the API functionality and on-board toolkits as quickly as possible}. Even before opening any toolkit's website, P3 shared, \textit{"first, I will see if there is any quick tutorial I can go through."} When exploring Fairlearn, this participant then proceeded directly to the ``Quickstart.'' Similarly, when exploring and comparing the two APIs, P4 and P8 revealed through their think-alouds that they were focusing on finding an existing notebook to follow so that they could begin using the toolkit as quickly as possible. 

While it is possible that participants were incentivized to find a quick solution in order to complete our study in the allotted time, when asked, practitioners suggested that they would do the same in their day to day work because, as P4 said, \textit{``there is always a time constraint in the real work.''} Our survey results further supported that same time pressure existed in practitioners’ daily work. Of twenty-four respondents who answered the final question of the survey, fourteen (58\%) agreed that they would look for the fastest available solution when encountering fairness issues in their work. We further expand upon this point in Section \ref{4.3.3}.

\subsubsection{Participants emphasized the importance of being able to integrate toolkits into their existing ML working pipelines}

Validating findings from Lee et al \cite{lee2020landscape}, most (ten out of eleven) participants in our interview study emphasized that, as a precondition for adoption in their workplaces, \textbf{fairness toolkits must be easily integrated into ML practitioners' existing workflows.} On this note, eight participants (P1, P2, P3, P4, P7, P8, P10, P11) all mentioned the resemblance of Fairlearn to scikit-learn \cite{scikit-learn} (an open-source, BSD-licensed machine learning python library widely used by ML community) in terms of API classes and functions. Since scikit-learn was their go-to library to build and evaluate ML models, participants believed that this similarity could help them incorporate toolkits into their current ML pipeline. To illustrate, when learning and comparing two toolkit APIs, P3 commented that \emph{``the first aspect that I am considering is how easily it can be directly concatenated in my current project.''} P3 appreciated that Fairlearn developers \emph{``introduce this tool with very standard scikit-learn [syntax]. It will have more people interested in using this tool because it’s already very aligned with what I’m comfortable using.''}

\subsubsection{Some participants wanted to use fairness toolkits as code repositories to build their own tool.} Lastly, a few participants expressed desires to \textbf{understand the implementation details behind methods in a fairness toolkit, as a starting point to build their own tools.} When asked how they might use Fairlearn or AIF360 in practice, P1 and P11 both entered toolkits' GitHub pages to see whether it would be possible to use the toolkits’ current implementations as references for implementing their own algorithms or toolkits. For instance, P11 needed domain-specific fairness assessment and mitigation methods for a recommendation system, which current toolkits did not include. Comparing the GitHub pages of two toolkits, P1 noted that AIF360’s GitHub \emph{``clearly listed out supported fairness metrics and mitigation algorithms.''} Similarly, P11 felt that AIF360 \emph{``has a more organized codebase to start with.''} Among the nineteen survey respondents who said they would rather build or extend their own tools, fourteen (79\%) of them selected the \emph{``need to understand the low-level implementation''} as one of their reasons for not using an out-of-the-box tool.

\subsubsection{\textbf{Implications}}
Our study surfaced that practitioners seek to use fairness toolkits for diverse purposes, leading to different expectations for what fairness toolkits should offer. Here, we suggest possible directions for fairness toolkits to support practitioners’ diverse needs. 

\begin{itemize}
    \item \textbf{Support practitioner learning within fairness toolkits}. Current toolkits are mainly designed to be used by practitioners for problem-solving \cite{richardson2021towards}. Our findings suggest that some practitioners might look to fairness toolkits as convenient sites to learn about unfamiliar ML fairness concepts. Future fairness toolkits might be explicitly designed as learning tools, for example with designated pages or interactive modules that introduce ML fairness concepts, procedures, and best practices. Toolkits might proactively direct practitioners to these pages both when they first begin using a toolkit, and at critical points throughout their use of a toolkit. Support for such ``just-in-time'' learning could help alleviate challenges identified in prior studies on fairness toolkits, in which practitioners struggled to pinpoint the resources they needed the most \cite{lee2020landscape, richardson2021towards}. 
    
    \item \textbf{Better support practitioners in  incorporating toolkits into their existing ML pipelines}. Echoing findings from Lee et al., our findings highlight the importance of supporting practitioners in more easily integrating fairness toolkits into their existing ML workflows \cite{lee2020landscape}. One possible strategy, noted by participants in our study, is to align syntax and function nomenclature used in fairness toolkits with that of existing popular programming languages (e.g., python~\cite{van1995python}, R~\cite{R}) and software libraries (e.g., scikit-learn~\cite{scikit-learn}, Pandas~\cite{mckinney2011pandas}, PyTorch~\cite{paszke2017automatic}, TensorFlow~\cite{tensorflow2015-whitepaper}) that are being widely used by data scientists and ML practitioners. 
    
    \item \textbf{Adapt to different time constraints and scaffold responsible use of fairness toolkits}. ML practitioners are often operating under time pressure, with minimal or no organizational processes in place to support fairness work, and with little to no training around ML fairness \cite{holstein2019improving, madaio2020co, madaio2021assessing, rakova2021responsible, metcalf2019owning, miceli2020between, miceli2022studying}.
    In light of these pressures and constraints, fairness toolkits face a difficult design challenge. They must be carefully designed to keep the time barrier to entry low enough that practitioners will be able to use these tools in their work, but without lowering the barrier to an extent that promotes irresponsible use. Approaches such as adding contextual interface warnings or including built-in checklists, discussed in \ref{4.1.4}, may be helpful in achieving this goal. In addition, for practitioners who desire detailed and thorough examples or extensible code, toolkits should provide well-contextualized ``worked examples'' \cite{richardson2021towards}, and aim to provide a flexible and clearly documented codebase. 

\end{itemize}

\subsection{Attempting to Use the Toolkits} \label{4.3}

In this section, we document findings from observing how practitioners actually attempt to assess and mitigate a fairness-related problem through concrete usage of the toolkits. In doing so, we aim to understand obstacles that may hinder effective use of fairness toolkits in practice.

\subsubsection{Participants desired better support for communication around fairness issues across different roles in an organization} Although current toolkits target individual developers, participants wanted to see \textbf{more functionality to support collaboration among multiple, diverse roles, including non-engineers}. After using the toolkits, we asked participants to explain their rationale for whether or not they would actually use a fairness toolkit in their own workflow.
Before delving into the patterns we observed, we note that, at the time of our interviews, the Fairlearn toolkit supported a dashboard which many participants saw in example notebooks and tutorial documentation. Although this widget is no longer being developed by the Fairlearn team, we observed many valuable takeaways from participants' use of this Jupyter notebook widget. Several participants (P1, P3, P6, P8, P9, P11) commented on the usefulness of the dashboard visualizations. For instance, before seeing the dashboard P6 said that the toolkit \emph{``was doing something to the data, but I can't really see what it's doing so that inherently makes me feel uncomfortable.''} Then, when presented with the Dashboard, they immediately felt more at ease with the idea of using Fairlearn. Along these lines, P3 and P6 noted that the dashboard allowed them to easily scan for potential fairness issues, which would be helpful under time constraints.

Participants connected the need for dashboard-like functionality to a broader \textbf{need for visualizations in order to explain fairness toolkits to non-engineering roles within an organization.} For example, P1 and P2 noted that the dashboard makes it simple to convey the toolkit's message to a larger audience and therefore motivate others to further engage with the toolkit. P2 and P9 reaffirmed this by claiming respectively that \emph{``a simple notebook format and compelling visualizations are needed for [organizational] leadership to adopt the toolkits''} and \emph{``from a business standpoint, the [Fairlearn] Dashboard is more helpful than Jupyter notebook.''} From our survey, when asked to what extent certain factors influence their decision on whether to perform fairness testing, of the twenty-four people who responded, all but two people (92\%) said ``stakeholder demand'' is important. This is supported by survey respondents who emphasized the role of external pressure, e.g. \emph{``if [concerns are] raised by superiors''} or \emph{“if there is a significant fairness issue, we would request more time and budget from the stakeholders.”} Another survey respondent noted that \emph{``often managers and software developers are faced with more complex realities''} and most out-of-box techniques that toolkits offer do not keep the \emph{``use cases, regulatory landscape, and real-world deployment in mind''}. Therefore, to encourage broader use, a respondent from our survey commented that they expected fairness toolkits to \emph{``provide relevant communication material (e.g. reports that can be circulated internally for the purpose of improving the software development process and/or scorecards that can be shared with external stakeholders like customers and investors).''}

\subsubsection{Participants desired more actionable guidance} Fairness-related nomenclature in toolkits' documentation was often unintuitive to participants. For example, regarding bias mitigation algorithms, P7 claimed that \emph{“[Fairlearn’s] CorrelationRemover is more intuitive as opposed to [AIF's] DisparateImpactRemover.”} Participants generally had a specific need in mind when delving into the documentation. P10 made the point that \emph{``toolkits need to do a better job of explaining [fairness metrics] and why users would want to use them''}. For instance, P9 explained that it's important to \emph{“situate users”} before delving into the intricacies of a fairness metric or bias mitigation algorithm. This is supported by a respondent from our survey, who said \emph{``I think most fairness toolkits are tailored for data scientists with overly complicated metrics. This makes it hard to explain the metrics to the business stakeholders which most of the time are looking for something that is intuitive and meaningful.''} As noted in the previous subsection, cross-functional communication is pivotal for fair ML efforts to have an impact in real organizational settings. To support this, we observed that participants often desired \textbf{more guidance and support in contextualizing toolkit functionalities or outputs, beyond the level of documentation provided by standard software packages.} 

\subsubsection{Participants frequently copied code directly from toy examples provided by the API} \label{4.3.3} Although participants mentioned the need for domain expert guidance in tailoring a fairness analysis to a particular problem, during the code writing stage, most of our participants \textbf{directly copy-pasted code that they found in tutorial notebooks or toolkit documentation}. Only \textit{one} participant (P7) attempted to reason through possible implications of their choices of particular fairness metrics or sensitive features. Although copy-pasting code from online examples is a common practice among developers when attempting to integrate a new software package into a working pipeline \cite{ko2011state}, doing so uncritically in the context of fair ML work may be particularly dangerous, given that fairness is highly context-dependent \cite{Dasch2020OpportunitiesFA, van2021effect, 10.1145/3178876.3186138, selbst2019fairness}.

\subsubsection{\textbf{Implications}} \label{4.3.4}
\begin{itemize}

    \item \textbf{Toolkits should provide use-specific and context-specific guidance.} Meaningful assessments are inherently contextual~\cite{cobbe2021reviewability}.
    Use-specific guidance is therefore important for helping practitioners place the toolkit's offerings within the broader context of concerns, while encouraging `cross-functional' collaborations where appropriate. For instance, a toolkit's documentation could indicate its relation to a specific legal or regulatory regime (e.g. GDPR, CCPA), which in turn can prompt interactions with colleagues in legal or compliance. Similarly, toolkits might contain context-specific guidance highlighting the considerations of various real-world applications. For example, toolkit developers could provide resources on the specific social and cultural contexts for a technical tutorial or example notebook. By doing so, toolkits could encourage practitioners to think more critically about the non-technical context in their approach to fairness-related issues. Offering context-specific guidance could also help practitioners avoid the issues with directly copy-pasting code from toolkits to their specific applications.
    
    \item \textbf{Opportunities for toolkits to support cross-functional collaboration and organizational buy-in.} Our results point to the need for toolkits to facilitate fairness related conversations and collaborations in a cross-functional setting, where not all team members will necessarily have much knowledge around either ML or ML fairness. It may be helpful, for example, for future toolkits to support such communication by generating visualizations and reports that are tailored for presentation to non-engineers. A large body of work in FAccT and HCI has surfaced how organizational factors (e.g., organizational cultures and incentives) impact ML fairness efforts in practice \cite{holstein2019improving, madaio2020co, madaio2021assessing, rakova2021responsible, metcalf2019owning, miceli2020between}. Explicitly designing toolkits for use by \textit{teams} and \textit{organizations}, not just engineers, may help to address some of the challenges practitioners face in getting organizational buy-in and pushing fairness work forward.

\section{Discussion and Future Directions}

\subsection{Fostering interdisciplinary communication and collaboration}

To better address fair ML issues, prior FAccT research has highlighted the need for contextualizing technical ML fairness work through cross-domain collaborations \cite{passi2018trust, madaio2020co} and bringing in the lived experiences of real-world stakeholders, especially those from marginalized communities \cite{holstein2019improving}. Our findings have shown that practitioners desire more support in contextualizing, communicating, and collaborating around ML fairness efforts, throughout the ML development lifecycle. From toolkit developers' perspectives, designing and creating a useful, context-specific example notebook also requires the help of domain experts. However, cross-functional collaboration in building potential solutions for ML fairness is intrinsically challenging due to background knowledge gaps \cite{lee2019webuildai, Yu2020KeepingDI}, conflicting values \cite{Srivastava2019MathematicalNV}, ambiguous goals \cite{Dasch2020OpportunitiesFA}, and organizational barriers \cite{madaio2021assessing, rakova2021responsible}. As current fairness toolkits are mainly developed by and for practitioners with technical backgrounds in ML and software development, it is our hope that future fairness toolkit developers will explore a broader possibility space for the role that toolkits could play in ML fairness practice. For example, fairness toolkits could be explicitly designed to foster communication and collaboration across diverse roles within an organization, or could support practitioners in connecting with other stakeholders with relevant domain knowledge or lived experiences outside their current organizations.

Designing toolkits % for 
to better facilitate
meaningful interdisciplinary communication and collaboration beyond technical fair ML work could also help practitioners avoid the "solutionism trap" discussed by Selbst et al., %referring 
which refers to the pitfalls of presuming a fairness-related issue can be solved through technical intervention alone \cite{selbst2019fairness}. 

Our think-aloud study engaged participants in learning about and using the fairness toolkits of today, which are currently designed to support relatively narrow, technical fair ML interventions.
As a result, we centred on a technical approach for tackling a socio-technical problem (educational resource allocation), without leaving much space for participants to critically question whether an ML approach is appropriate in the first place.
In  practice, it is important to consider a broader space of potential remedies to fairness concerns.
Future fairness toolkits might be designed to scaffold developers through broader reflection processes, for example by prompting them to ask "should we build ML in the first place?" or "could we solve this fairness issue solely through technical intervention?".

In Section \ref{4.1.4}, we briefly discussed opportunities for community building, building upon the social channels (e.g., community Slack or Discord channels) that existing toolkits already provide. Future work is needed to explore ways to build sustainable, diverse, interdisciplinary communities of practice. Toolkit developers interested in pursuing this vision may be able to draw lessons from other open-source community-building efforts \cite{halfaker2013rise, Bryant2005BecomingWT}, or from recent work exploring ways to support collective algorithm auditing \cite{pmlr-v81-buolamwini18a,ali2019discrimination,ShenEveryDayAuditing:2021, cabrera2021discovering}. In short, we advocate for future fairness toolkits to position themselves as socio-technical systems that enable more collaborative approaches to ML fairness practice. 

\subsection{Future directions for evaluating fairness toolkits}

With more fairness toolkits being developed and deployed by research institutions and private companies, how might we support more effective and responsible use of fairness toolkits in the future? Beyond the design implications presented above, we hope that our work will inspire future empirical evaluations of practitioners' use of fairness toolkits, to empirically inform toolkit designs. Future work should aim to engage practitioners from more diverse regions \cite{sambasivan2021re}, domains \cite{passi2019problem, wang2020factors, holstein2019improving}, and organizational contexts \cite{passi2018trust, madaio2020co}, given that practices and challenges may vary significantly across each of these dimensions \cite{kleinberg2018inherent, madaio2021assessing}.

As we observed in our study, with little to no training around ML fairness, several practitioners fell prey to the ``fairness through unawareness'' \cite{dwork2012fairness} trap (Section \ref{4.1.1}). Moreover, practitioners could seek for the most convenient solutions, rather than the appropriate ones, with minimal or no organizational processes in place to support fairness work and workplace time constraint(Section \ref{4.1.3}, Section \ref{4.3.3}), failed to consider the procedural, contextual nature of ML fairness \cite{passi2019problem, mulligan2019thing, mitchell2018prediction, wang2020factors, selbst2019fairness}. Future toolkit designers should prevent practitioners from committing to other common ML fairness pitfalls discussed by prior research, for example, ``fairness gerrymandering,'' \cite{kearns2018preventing} ``solutionism trap,'' \cite{selbst2019fairness} and ``formalism trap'' \cite{selbst2019fairness}. To this end, toolkit designers should empirically study how practitioners apply fairness toolkits to more complex tasks with a wider range of datasets. 

Our study investigates individual practitioners' use of fairness toolkits. However, as discussed in Section \ref{4.3} when tackling complex, multi-faceted fairness issues in real-world settings, fairness toolkits need to support interactions and collaborations across diverse roles, including non-engineers (Section \ref{4.3.4}). Future work is needed to explore how \emph{teams} in organizations \emph{collectively} use fairness toolkits on real-world tasks. This could enable insight into the ways team dynamics might add additional frictions to toolkit use \cite{salehi2015we}. Only through longer-term ethnographic studies on in-house use of fairness toolkits can toolkit developers fully understand toolkits' uses, limitations, and potential impacts. Finally, we encourage fairness toolkit developers and researchers to not only use findings from such studies to iterate on toolkits' designs, but also to publish key empirical findings in the format of white papers, blog posts, and toolkit documentation, in the interest of communicating toolkit usage patterns and anti-patterns (cf. \cite{guo2021ten}).

\section{Conclusion}

In this study, we conducted the first empirical exploration of how industry practitioners (try to) work with fairness toolkits in practice. Through our think-aloud methods and accompanying anonymous survey, we found that practitioners needed support from toolkits in order to help them better contextualize fairness issues, as well as to assist them in fostering communication and collaboration with non-technical peers in their organizational settings. Additionally, we discovered numerous design implications for future developers of toolkits that seek to address complex, socio-technical problems. Beyond this, we hope our findings provide guidance for the creation of interdisciplinary communities, dedicated to providing a holistic space in order to combat fairness-related issues.

\begin{acks}

We thank all industry practitioners who signed up for our interview study, replied to our recruitment emails, and finished the think-aloud study. We also thank anonymous industry practitioners who filled out our online survey. We would like to express our gratitude to Michael Madaio, Miro Dudik, Roman Lutz, Hilde Weerts, and Alex Cabrera for their feedback on different stages of this work. Finally, we thank our anonymous reviewers for their thoughtful feedback that help us further improve this work. This work was supported by the National Science Foundation (NSF) under Award No. IIS-2001851, CNS-1952085, IIS-2000782, the NSF Program on Fairness in AI in collaboration with Amazon under Award No. IIS-1939606, the award from Jacob Foundation for CERES network, the Carnegie Mellon University Block Center for Technology and Society Award No. 53680.1.5007718, Aviva and the UK Engineering and Physical Science Research Council (EP/R033501/1, EP/P024394/1).

\end{acks}

\end{itemize}

\clearpage
\bibliographystyle{ACM-Reference-Format}
\bibliography{citation}

\end{document}